# A time-dependent scoring system for soccer leagues.

**Manuel B. Cruz (LEMA-ISEP/IPP, Portugal)** | mbc@isep.ipp.pt

**Miguel Pinho (Vanguard Method, Portugal)** | miguel.pinho@vanguardmethod.pt

**Sandra Ramos (LEMA-ISEP/IPP, Portugal)** | sfr@isep.ipp.pt

## Abstract:

In this paper, a new continuous scoring system for soccer is proposed, based on the proportion of time that a team is winning, losing or tied. Several simulations are made applying this technique to complete seasons of different leagues. As some preliminary fundamental analysis previews, the simulation results show that this proposal compacts the gap between several teams, increasing the competitiveness of the championships and making the matches even more challenging as every minute may count in order to get the best rank in the final standings. Based on those results, some characteristics of the model are highlighted and a discussion on some of the advantages, disadvantages and practical issues regarding a hypothetical implementation of this model is made. Finally, some ideas to tackle the less positive effects of an implementation in real leagues is also provided.

**Keywords:** Scoring System, Soccer, Competitiveness, La Liga, Premier League, Liga NOS

## 1. Introduction

A soccer game is one of the most enthusiastic subjects in the world. Its importance may be, for example, measured throughout the number of people directly connected with the sport, that the FIFA report of 2007 (FIFA, 2007) estimate in *"265 million male and female players and a further five million referees, coaches and other officials, a grand total of 270 million people, or four per cent of the world's population"*.

Also, the online presence is huge, with the top five team's websites in Europe having something like 38.3 million visitors online per month, according to UEFA (Daily Mail, 2017), while the aggregate revenue for the top 20 Money League clubs rose 6% in 2016-2017 to €7.9 billion (Boor, Hanson, & Ross, 2018). One of the main aspects for this numbers is related with the spectators, attending live or through TV broadcast which reaches outstanding numbers. According to (Hellier, Penty, & Mayes, 2017) only in Premier League, the 2017-2018 early season was attended through the two companies with the rights for the most games (Sky and BT) by 1.5 million viewers per match. Several aspects are in constant changing throughout this sport. The management, the tactics, the technological resources made available to managers, coaches, referees or public, had suffered several updates in the last years. Despite this, the scoring system is the same for several years although some proposals had been made mainly related with game theory and/or probabilistic points of view (Haugen, 2007; Fernandez-Cantelli & Meeden, 2003). In fact, since 1995 all the main leagues and championships use the 3-points-for-a-win in continuous competitions like national leagues or group stages at world championships, making this trinary scoring system (defeat, draw or win) the standard in soccer.

This system was introduced to promote attacking soccer. In fact, make a win yielded more two points than a draw should leads to goals galore. However, the effect of the 3-points-for-a-win system still up for debate, since the value of a win makes "parking the bus" at 1-0 a more desirable option (reference).

As generally accepted, attack playing is a joy to watch and even more of a joy to play, so scoring systems that encourage more attacking play since the early beginning of a match will certain benefit this sport attractiveness.

This paper is organized as follows: in Section 2, we propose a new scoring system based on the percentage of time that a team is (or is not) in advantage through its opponent during the game length. This system is applied to real results of Premier League, La Liga which are analysed with some detailed in Section 3. On Section 4, a global analysis is made as well as the discussion of the proposal merits and possible implementation problems in the real world. Finally, for these questions some possible updates are proposed in order to left some ideas for future works on this area.

## 2. The continuous time-dependent scoring system

The scoring system presented on the following lines, relies on simple premise: the final points attributed to each team should be proportional to the length of time where each team was winning/tied/losing against its opponent with respect to the full length of the match. In general, the total points earned by a team *A* when playing a match against an opponent *B*, may be calculate through:

$$P_A = \frac{\alpha_w T_{wA} + \alpha_d T_{dA} + \alpha_l T_{lA}}{T_m}$$

where:

$P_i$ - final points won by team *i, i=A,B*

$\alpha_w$ - weight assigned to the wining periods

$\alpha_d$ - weight assigned to the draw periods

$\alpha_l$ - weight assigned to the losing periods

$T_{WA}$ - Time quantity for which team *i* was winning in the match, *i=A,B*.

$T_{di}$ - Time quantity for which the match was draw, *i=A,B*.

$T_{li}$ - Time quantity for which team *i* was losing in the match, *i=A,B*.

$T_m$ - Full-length of the match (including extra-time).

In this formulation, it is straightforward that $T_{WA} + T_{dA} + T_{lA} = T_m$, $T_{WA} = T_{lB}$ and that $T_{dA} = T_{dB}$. As so, the sum of points given to a set of two teams playing against each other in a single match will become:

$$P_A + P_B = \frac{\alpha_W T_{WA} + \alpha_d T_{dA} + \alpha_l T_{lA} + \alpha_W T_{WB} + \alpha_d T_{dB} + \alpha_l T_{lB}}{T_m}$$

$$P_A + P_B = \frac{(\alpha_W + \alpha_l)T_{WA} + 2\alpha_d T_{dA} + (\alpha_W + \alpha_l)T_{lA}}{T_m}$$

$$P_A + P_B = \frac{(\alpha_W + \alpha_l)(T_{WA} + T_{lA}) + 2\alpha_d T_{dA}}{T_m}$$

$$P_A + P_B = \frac{(\alpha_W + \alpha_l)(T_m - T_{dA}) + 2\alpha_d T_{dA}}{T_m}$$

$$P_A + P_B = (\alpha_W + \alpha_l) + \frac{(2\alpha_d - \alpha_W - \alpha_l)T_{dA}}{T_m}$$

In the forthcoming simulations, we will assign the values (3,1,0) to the $(\alpha_w, \alpha_d, \alpha_l)$ previously defined weights as they derived from the actual scoring system in major soccer leagues and championships. In this case, the last equation becomes

$$P_A + P_B = 3 - \frac{T_{dA}}{T_m}$$

From this expression, it is straightforward that the sum of points assigned to both teams on the end of the match is a continuous value in the interval [2,3[ (right opened as a goal takes, at least, a few seconds to score), while each team may obtain a value belonging to the interval ]0,3[. This may be checked in Figure 1, that represents the empirical cumulative distribution function plotted with the real data from La Liga 2015-2016. Through this representation it may be seen that, accordingly to the dot curve, in around 40% of the games, the competing teams got 0 points, in a little more than 20% they got 1 point and in almost 40% they got 3 points at the end of the match. For the same season, and if the new system had been used, the blue curve shows that in around 25% of the matches, one team gets 0.5 points or less, 45% with less than 1 point, 75% with less than 2 points and in the remaining 25% the team gets between 2 and 3 points. Nowadays, in the 3-for-a-win scoring system, the sum of the points gained by both teams belongs to the discrete set {2,3} and each team may get a value of within the {0,1,3} set. As consequence, the final values obtained by teams will be smaller than the ones that are attributed nowadays, implying that the gap between teams also will tend to become smaller, increasing the championship competitiveness.

Another important aspect is that each minute of extra time ($ET$) may give at most, $2*1/(90 + ET)$ points, in opposition to the presently adopted scoring system where, for example, one goal at last minute may originate two extra points for the scoring team. Additionally, in the proposed system, the process is symmetric in the sense that a team winning for the first $T_m$ minutes and tie for the remaining 90-$T_m$, will get the same number of points as if the game is tied for all the game and it scored the victory goal at minute 90-$T_m$.

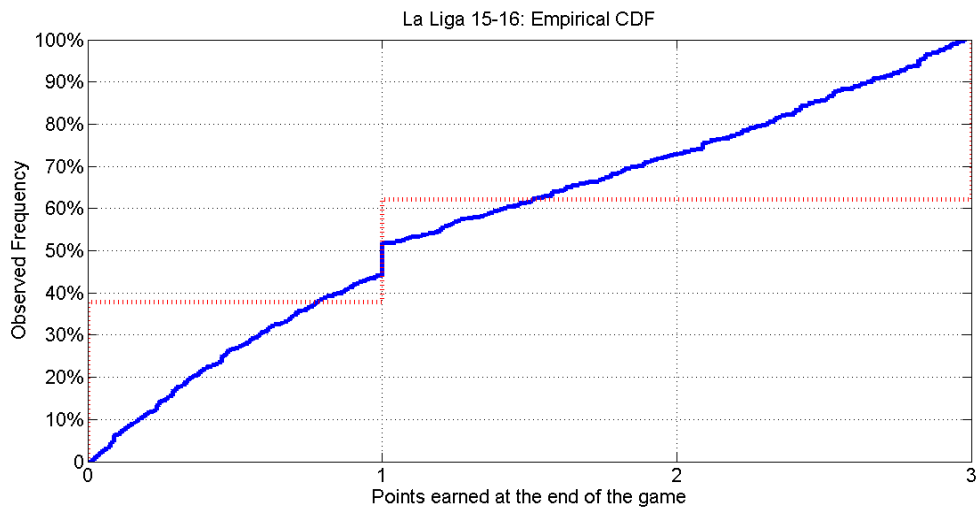

*Figure 1: Empirical cumulative distribution function. Number of points earned per team with new system (full line) and 3-for-a-win (dotted line) when applied to La Liga 15-16 matches.*

Also, a team winning for a period of 30 minutes (1/3 of the game length) and losing for the other 60' will get the same points as if the games end with 0-0. All of these, will certainly induce new management paradigms on defining the tactics for a match, as well as would need new regulations with respect to the time variable. Indeed, extra time, time discounts regarding substitutions and players assistance, should be regulated as they will have considerable importance on match points.

## 3. Simulations with real data.

In order to check the numerical outcomes of the proposed scoring system, we simulated the proposal within the games for four different seasons: Premier league (England), La Liga (Spain) and Liga NOS (Portugal) for both 2015-2016 and 2016-2017 seasons. For every game of each League season, the timing of each scored goal was registered. This information is available through several websites as, for example, Premier League (Premier League, 2016), La Liga (La Liga, 2016), Liga NOS (Liga Portugal, 2017) official websites, or even through the more general website ZeroZero.pt (ZOS, Lda, 2016). As those official results are presented in minutes, the following calculations have an absolute error smaller than 2*59/(90*60)≈0.02 points per goal if available data is obtained by truncation, or around 0.01 if the goal timing is rounded by the websites sources for the nearest minute. Also, and because we didn't found any reliable source for the exact match length, we consider by default that all the games last for 90 minutes, except for the ones where the last goal was scored beyond that time. In this case, we consider the length of the game, $T_M$, as that moment. As our main concern is in the discussion of hypothetical merits of this approach and not within the results themselves, those errors don't seem of upmost importance in this paper context.

In the next tables, we present the final standings of each championship based on our scoring proposal (denoted as "New system") and compare them with the official 3-for-a-win results. As we are proposing a methodology that changes the final standings scale, next to the "Final Points" columns we also present the % of final points that each team reached with respect to the points of the team that stands in the first place. The "Min to upper pos" last column presents, for each team, the number of minutes that would be needed to anticipate a victory goal in order to that team overtake the precedent opponent in the final standings. Following each table, we also present the representation of each team position across the entire season for both new system (top plot) and 3-for-a-win (bottom plot). Finally, a figure representing the top-4 and bottom-4 teams in each of the scoring systems is also shown, as it may provide a detailed insight on the championship evolution regarding the battles for some positions.

## 3.1. Premier League 2015-2016

In 2015-2016, Premier league had 20 teams, that played 760 matches between themselves. In the 3-for-a-win scoring system, the mean of points per match was 1033/760=1.36 points per game. During the whole season, there were four different teams on leadership (Manchester City, Manchester United, Leicester and Arsenal), corresponding to 5 changes on the leadership. If the new system had been applied, the mean points per match dropped to 1.25 points/game and there were six teams on leadership (Manchester City, Leicester, West Ham, Manchester United, and Arsenal), that swapped 15 times on the first position between themselves. Leicester City, the 2015-2016 champion, finished with an advantage of 10 points for the second placed (Arsenal). To eliminate this gap, Arsenal would swap 5 draws into wins, in order to be tied with Leicester, according to the present scoring rules. Using our time dependent system proposal, that gap became shorter with Leicester winning only by 0.73 points to Arsenal. This gap may be translated into terms of how much time a goal that gave a victory to the second classified should take place to close that gap. In this case, if Arsenal scored, in only one game, the victory goal 33 minutes earlier, it would be the champion of Premier League 2015-2016. In fact, at least half of the teams, only had to anticipate one victory goal 55 minutes earlier in order to overtake the competitor that finished the league in the upward position.

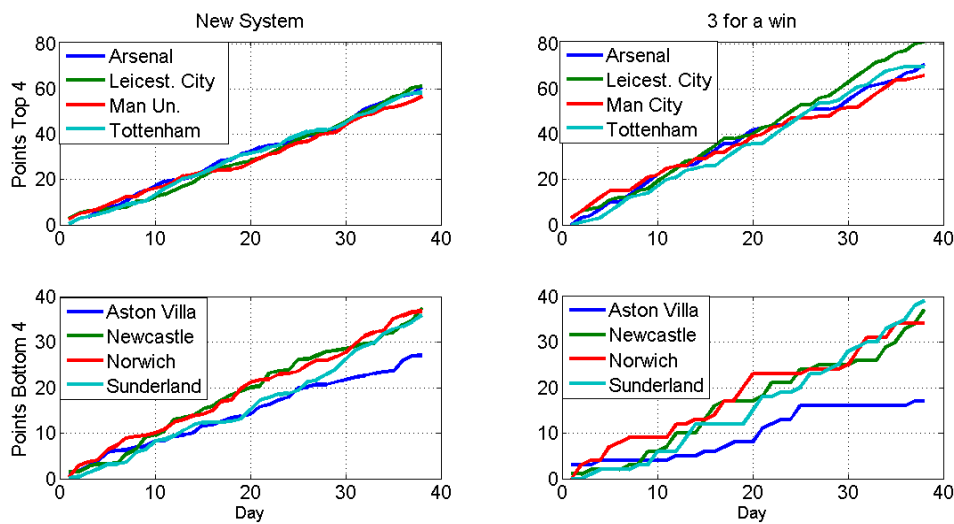

*Figure 2: Premier League 2015/2016 - Details of top and bottom standings*

Regarding the competitiveness between the top four teams, beside their final points stands in a narrower band (>90%) as seen in Table 1, that behaviour is also observed all across the season, as it may be seen on the top plots of Figure 2.

| | **New System** | | | **3-for-a-win** | | | **Draw->Win** |
|---|---|---|---|---|---|---|---|
| Rank | Team | Final Points | % pts of 1st | % pts of 1st | Final Points | Team | Rank | Min to upper pos |
| 1 | Leicester | 61,41 | 100 | 100 | 81 | Leicester | 1 | |
| 2 | Arsenal | 60,68 | 99 | 88 | 71 | Arsenal | 2 | 33 |
| 3 | Tottenham | 58,48 | 95 | 86 | 70 | Tottenham | 3 | 99 |
| 4 | Man. United | 56,68 | 92 | 81 | 66 | Man. City | 4 | 81 |
| 5 | Southampton | 55,46 | 90 | 81 | 66 | Man. United | 5 | 55 |
| 6 | Man. City | 53,66 | 87 | 78 | 63 | Southampton | 6 | 81 |
| 7 | Chelsea | 52,84 | 86 | 77 | 62 | West Ham | 7 | 37 |
| 8 | West Ham | 52,28 | 85 | 74 | 60 | Liverpool | 8 | 25 |
| 9 | Liverpool | 51,65 | 84 | 63 | 51 | Stoke City | 9 | 28 |
| 10 | Stoke City | 48,34 | 79 | 62 | 50 | Chelsea | 10 | 149 |
| 11 | Everton | 48,04 | 78 | 58 | 47 | Everton | 11 | 13 |
| 12 | Swansea | 45,24 | 74 | 58 | 47 | Swansea | 12 | 126 |
| 13 | Watford | 44,30 | 72 | 56 | 45 | Watford | 13 | 43 |
| 14 | W. Bromwich | 41,49 | 68 | 53 | 43 | W. Bromwich | 14 | 126 |
| 15 | Bournemouth | 41,24 | 67 | 52 | 42 | Crystal Palace | 15 | 11 |
| 16 | Crystal Palace | 38,93 | 63 | 52 | 42 | Bournemouth | 16 | 104 |
| 17 | Newcastle | 37,41 | 61 | 48 | 39 | Sunderland | 17 | 68 |
| 18 | Norwich | 36,77 | 60 | 46 | 37 | Newcastle | 18 | 29 |
| 19 | Sunderland | 35,80 | 58 | 42 | 34 | Norwich | 19 | 44 |
| 20 | Aston Villa | 26,99 | 44 | 21 | 17 | Aston Villa | 20 | 396 |

*Table 1: Comparison on Premier League 2015-2016 final standings.*

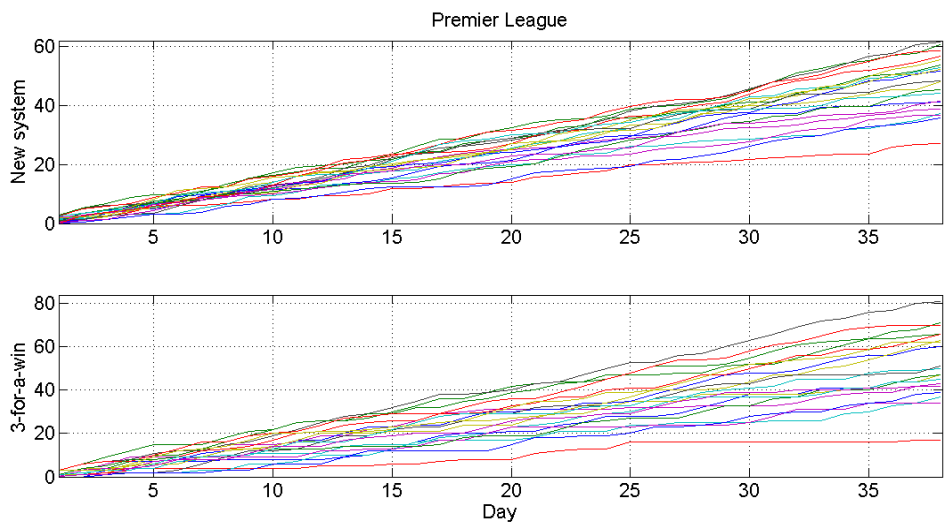

*Figure 3: Premier League 2015/2016 - Standings across the season*

The same is also verified when analysing the bottom 4 teams, as they stand on a 44%-61% band regarding Leicester final points with the new system, against the 21%-48% obtained with the traditional 3-for-a-win. It is also important to notice that the new system would had relied Sunderland to the second league instead of Newcastle United. As curiosity, Sunderland finished the 2016/2017 League in the last position, and Newcastle United was promoted again to Premier league on the end of that season.

### 3.2. La Liga 2015-2016

In 2015-2016, the Spanish League had 20 competing teams. Regarding the 760 different matches played, the mean of points per match was 1048/760=1.38 points per game. During the whole season, there were 5 different teams to reach the top of the table (Eibar, Celta de Vigo, Barcelona, Real Madrid and Villareal), corresponding to 6 changes on the leadership. With the proposed changes the mean points per match was reduced to 1.27 points/game while there would be 5 teams on leadership (Espanyol, Real Madrid, Celta de Vigo, Eibar and Barcelona), that swap 8 times on the first position between themselves.

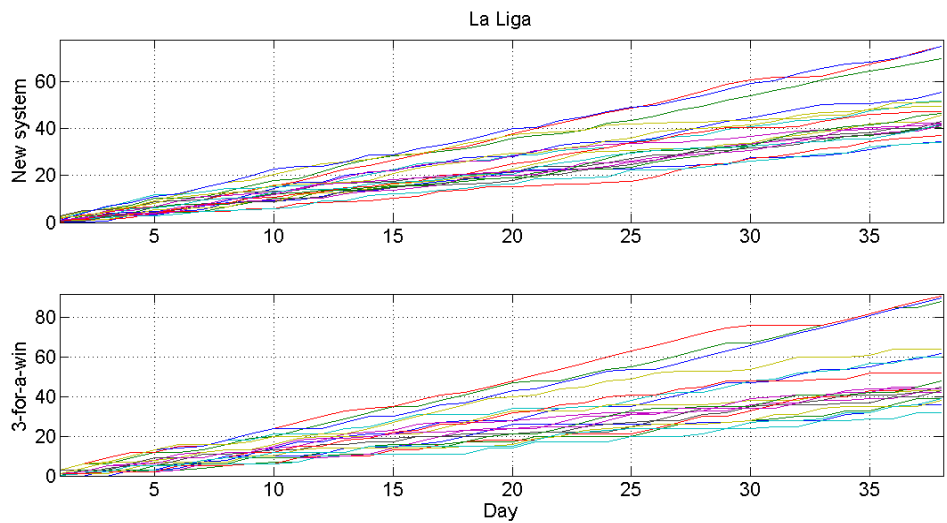

*Figure 4: La Liga 2015/2016 - Standings across the season*

This was one of the most competitive leagues ever, as resumed in Table 2 and plotted on Figures 4-5, with the champion Barcelona, winning the League only one point above the second classified Real Madrid.

With the time dependent scoring system, that gap became only 0.02 points disregarding the eventual inaccuracies with respect to rounding the exact score moment. This mean that, if in just one of the Real Madrid victories he had scored only 0.7 minutes (around 40 seconds) before, he would get the championship title. As our dataset precision was in minutes, we don't have enough accuracy to decide who would be the champion on this season.

At least half of the teams, only had to score the victory goal to 34 minutes early in order to overtake the competitor that finished the league in the upward position.

There is also a huge difference regarding the classifications of Rayo Vallecano and Las Palmas, as their rank swaps from 11th->18th (Las Palmas) and 18th->10th (Rayo Vallecano).

| | **New System** | | | **3-for-a-win** | | | | **Draw->Win** |
|---|---|---|---|---|---|---|---|---|
| Rank | Team | Final Points | % pts of 1st | % pts of 1st | Final Points | Team | Rank | Min to upper pos |
| 1* | Barcelona | 75,02 | 100 | 100 | 91 | Barcelona | 1 | |
| 2* | Real Madrid | 75,00 | 100 | 99 | 90 | Real Madrid | 2 | 0,7* |
| 3 | At. Madrid | 69,78 | 93 | 97 | 88 | At. Madrid | 3 | 234,7 |
| 4 | Athletic Club | 55,61 | 74 | 70 | 64 | Villarreal | 4 | 637,8 |
| 5 | Celta de Vigo | 51,83 | 69 | 68 | 62 | Athletic Club | 5 | 170,3 |
| 6 | Villarreal | 51,31 | 68 | 66 | 60 | Celta de Vigo | 6 | 23,0 |
| 7 | Eibar | 49,55 | 66 | 57 | 52 | Sevilla | 7 | 79,3 |
| 8 | Sevilla | 47,36 | 63 | 53 | 48 | Málaga | 8 | 98,9 |
| 9 | Real Sociedad | 46,37 | 62 | 53 | 48 | Real Sociedad | 9 | 44,3 |
| 10 | R. Vallecano | 45,62 | 61 | 49 | 45 | Real Betis | 10 | 34,3 |
| 11 | Espanyol | 43,02 | 57 | 48 | 44 | Las Palmas | 11 | 116,7 |
| 12 | Dep. Coruña | 42,58 | 57 | 48 | 44 | Valencia | 12 | 20,1 |
| 13 | Sporting Gijón | 42,14 | 56 | 47 | 43 | Espanyol | 13 | 19,7 |
| 14 | Valencia | 42,11 | 56 | 47 | 43 | Eibar | 14 | 1,3 |
| 15 | Real Betis | 41,53 | 55 | 46 | 42 | Dep. Coruña | 15 | 26,1 |
| 16 | Málaga | 41,16 | 55 | 43 | 39 | Granada | 16 | 16,7 |
| 17 | Granada | 40,41 | 54 | 43 | 39 | Sp. Gijón | 17 | 33,7 |
| 18 | Las Palmas | 37,01 | 49 | 42 | 38 | R. Vallecano | 18 | 153,1 |
| 19 | Getafe | 34,36 | 46 | 40 | 36 | Getafe | 19 | 119,1 |
| 20 | Levante | 33,96 | 45 | 35 | 32 | Levante | 20 | 17,8 |

*Table 2: Comparison on La Liga 2015-2016 final standings. (*) There is not enough precision on time data to evaluate the rank position.*

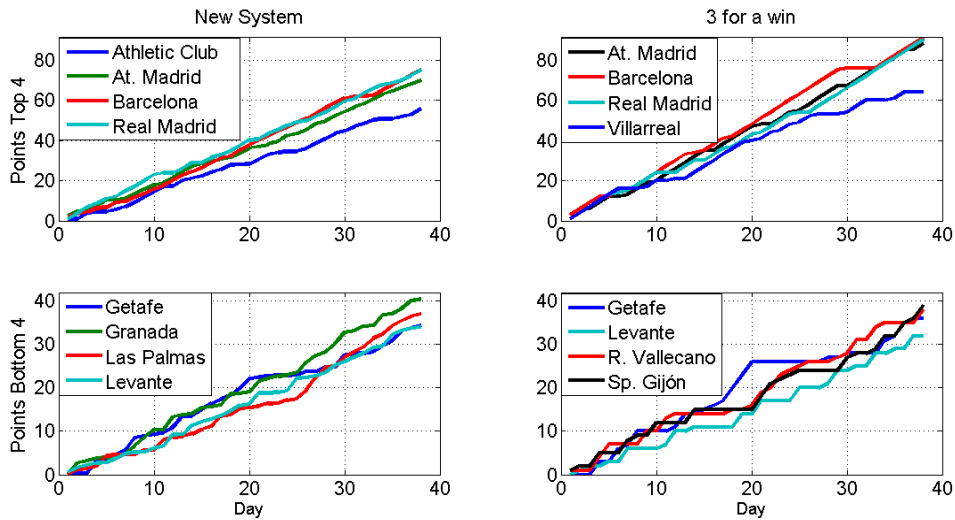

*Figure 5: La Liga 2015/2016 - Details of top and bottom standings*

### 3.3. Liga NOS 2015-2016

The Portuguese League has 18 competing teams since 2014-2015. This number of teams implies that in each season, there are 612 different matches played. In 2015-2016, the mean of points per match was 842/612=1.38 points per game. The new system would decrease this value to 1.26 points per game, and the 4 different teams on the leadership (Porto, Setúbal, Sporting and Benfica) had swap positions for 11 times. Officially, although the nominal list was slightly different (Benfica, Arouca, Porto and Sporting), there were also 4 leaders for this season who have swapped positions for 8 times.

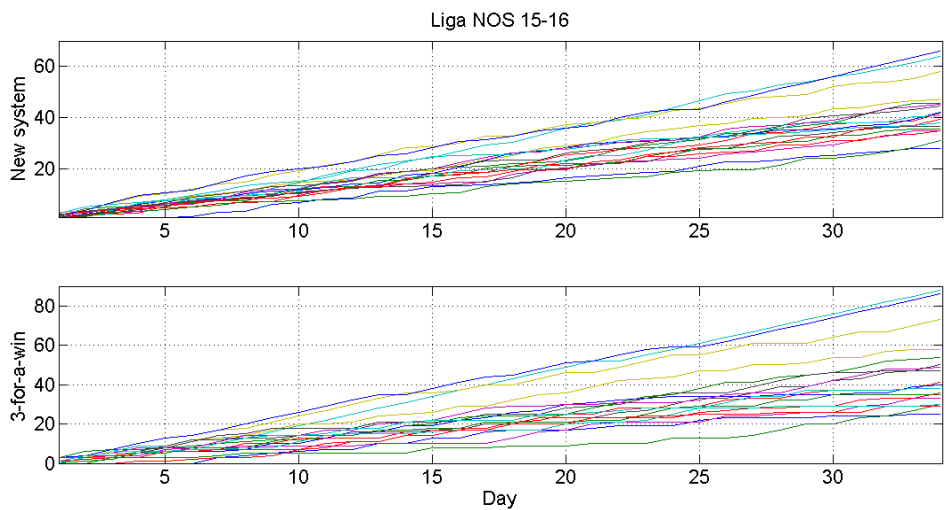

*Figure 6: Liga NOS 2015/2016 - Standings across the season*

|      | **New System** |                |              | **3-for-a-win** |                |              |      | **Draw->Win** |
|------|----------------|----------------|--------------|-----------------|----------------|--------------|------|---------------|
| Rank | Team           | Final Points   | % pts of 1st | % pts of 1st    | Final Points   | Team         | Rank | Min to upper pos |
| 1    | Sporting       | 66,2           | 100          | 100             | 88             | Benfica      | 1    |               |
| 2    | Benfica        | 63,8           | 96           | 98              | 86             | Sporting     | 2    | 106           |
| 3    | Porto          | 56,8           | 86           | 83              | 73             | Porto        | 3    | 318           |
| 4    | Braga          | 46,9           | 71           | 66              | 58             | Braga        | 4    | 442           |
| 5    | Arouca         | 45,6           | 69           | 61              | 54             | Arouca       | 5    | 60            |
| 6    | P. Ferreira    | 45,2           | 68           | 57              | 50             | Rio Ave      | 6    | 19            |
| 7    | Rio Ave        | 44,6           | 67           | 56              | 49             | Paços Ferreira | 7  | 26            |
| 8    | Guimarães      | 42,1           | 64           | 53              | 47             | Estoril      | 8    | 114           |
| 9    | Belenenses     | 41,7           | 63           | 47              | 41             | Belenenses   | 9    | 20            |
| 10   | Nacional       | 40,6           | 61           | 45              | 40             | Guimarães    | 10   | 47            |
| 11   | Moreirense     | 39,4           | 60           | 43              | 38             | Nacional     | 11   | 55            |
| 12   | Setúbal        | 37,9           | 57           | 41              | 36             | Moreirense   | 12   | 65            |
| 13   | Estoril        | 36,7           | 55           | 40              | 35             | Marítimo     | 13   | 56            |
| 14   | Marítimo       | 35,6           | 54           | 38              | 33             | Boavista     | 14   | 51            |
| 15   | Boavista       | 35,0           | 53           | 34              | 30             | Setúbal      | 15   | 27            |
| 16   | Un. Madeira    | 34,3           | 52           | 34              | 30             | Tondela      | 16   | 30            |
| 17   | Tondela        | 30,9           | 47           | 33              | 29             | Un. Madeira  | 17   | 155           |
| 18   | Académica      | 27,8           | 42           | 28              | 25             | Académica    | 18   | 136           |

Table 3: Comparison on Liga NOS 2015-2016 final standings.

In this season, if the new methodology had been applied, there would a swap between the first and second final standings, as Sporting would be champion instead of Benfica. The 3rd place was taken by Porto in any of the scoring options in both cases at a considerable distance from the 2nd place. Although in different positions, the last 3 ranked clubs are the same in both methodologies and the gaps between them is almost the same in terms of the percentage of points earned with respect to the season champion. Nevertheless, the competitiveness of this teams across the season is higher on the new system, as it may be seen in Figure 6. This is confirmed in a more broader sense, as the overall changes in the ranks across the same season increases more than 10% as it may be seen in **Erro! A origem da referência não foi encontrada.**

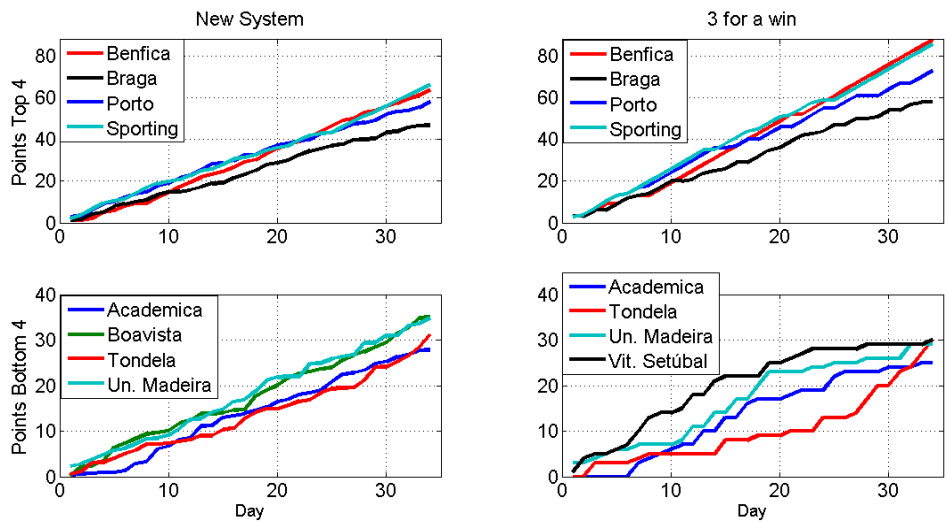

*Figure 7: Liga NOS 2015/2016 - Details of top and bottom standings*

### 3.4. Liga NOS 2016-2017

We also analysed the 2016-2017 season for the Portuguese league, where there was an average of 838/612=1.37 points per game. If the new system had been applied, that value would decreased to 1.26 points that is on the range of values defined by all the league seasons analysed previously. Benfica would be the champion independently of which scoring system is applied, but within the new system the margin to the $2^{nd}$ placed team is narrowed (from 93 to 99% of the points). In this scenario, Boavista, Chaves, Sporting and Benfica would had reach the first place somewhere through the season, giving place to 9 swaps on that position. Tondela, who had finish in the $17^{th}$ official position, would became $12^{th}$, in a very comfortable position. This because, in several games, they were draw or in advantage for a big part of several games but they consented a decisive goal in the few minutes of the match. For example, they lost home games at 81' (Belenenses), 87' (Moreirense), 82' (Feirense) and allowed a draw at 89' (Boavista). When visiting its opponents, they consent the defeat goal at 90' (Nacional), and draws at 95' (Sporting) or at 84' (Moreirense). Regarding the official results, only 3 teams reached the top (Porto, Sporting and Benfica) with only two changes in the leadership.

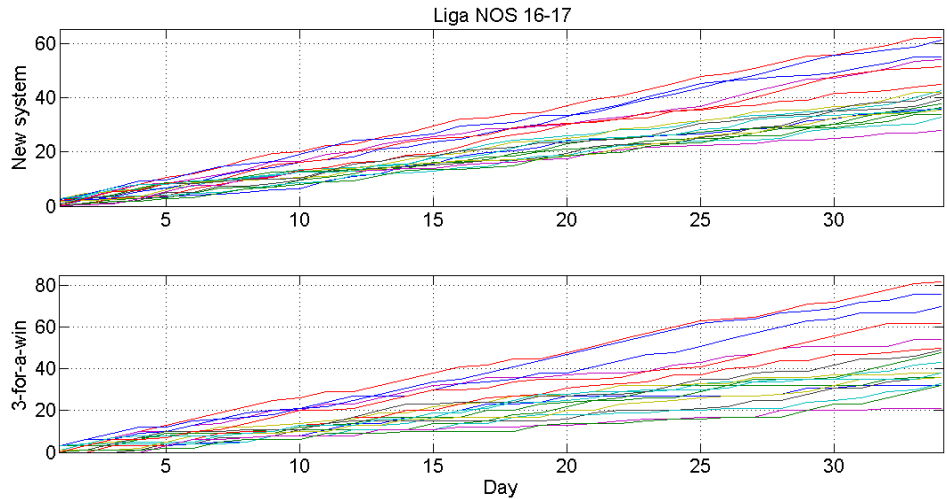

*Figure 8: Liga NOS 2016/2017 - Standings across the season*

| | New System | | | 3-for-a-win | | | | Draw->Win |
|---|---|---|---|---|---|---|---|---|
| Rank | Team | Final Points | % pts of 1st | % pts of 1st | Final Points | Team | Rank | Min to upper pos |
| 1 | Benfica | 62,1 | 100 | 100 | 82 | Benfica | 1 | |
| 2 | Sporting | 61,4 | 99 | 93 | 76 | FC Porto | 2 | 30 |
| 3 | FC Porto | 55,2 | 89 | 85 | 70 | Sporting | 3 | 280 |
| 4 | Braga | 53,9 | 87 | 76 | 62 | Guimarães | 4 | 56 |
| 5 | Guimarães | 51,5 | 83 | 66 | 54 | Braga | 5 | 109 |
| 6 | Marítimo | 44,9 | 72 | 61 | 50 | Marítimo | 6 | 296 |
| 7 | Boavista | 42,5 | 69 | 60 | 49 | Rio Ave | 7 | 108 |
| 8 | Chaves | 42,0 | 68 | 59 | 48 | Feirense | 8 | 24 |
| 9 | Rio Ave | 41,5 | 67 | 52 | 43 | Boavista | 9 | 20 |
| 10 | Estoril | 39,3 | 63 | 46 | 38 | Chaves | 10 | 101 |
| 11 | Feirense | 37,7 | 61 | 46 | 38 | Estoril | 11 | 72 |
| 12 | Tondela | 36,5 | 59 | 46 | 38 | Setúbal | 12 | 55 |
| 13 | Setúbal | 36,0 | 58 | 44 | 36 | Belenenses | 13 | 23 |
| 14 | Arouca | 35,7 | 58 | 44 | 36 | P. Ferreira | 14 | 10 |
| 15 | P. Ferreira | 35,6 | 57 | 40 | 33 | Moreirense | 15 | 8 |
| 16 | Belenenses | 33,8 | 54 | 39 | 32 | Arouca | 16 | 82 |
| 17 | Moreirense | 32,7 | 53 | 39 | 32 | Tondela | 17 | 46 |
| 18 | Nacional | 27,8 | 45 | 26 | 21 | Nacional | 18 | 224 |

Table 4: Comparison on Liga NOS 2016-2017 final standings.

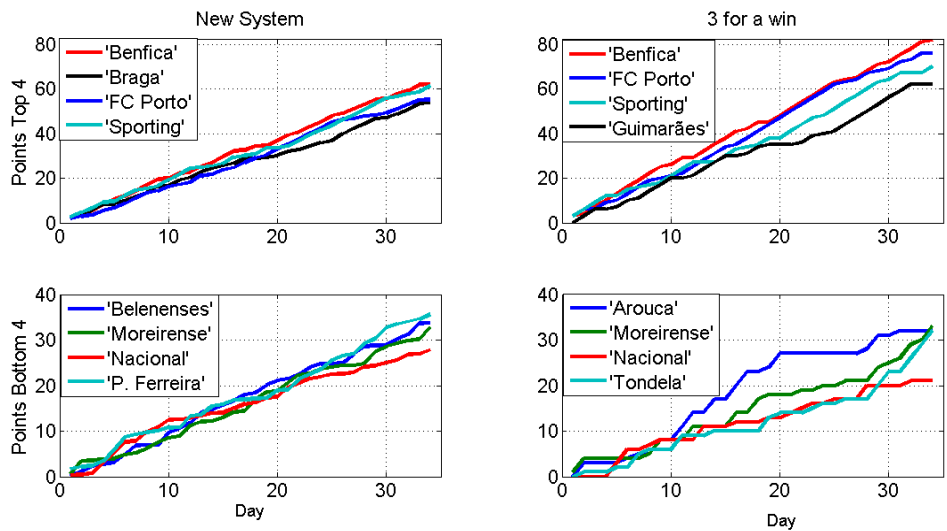

*Figure 9: Liga NOS 2016/2017 - Details of top and bottom standings*

## 4. Global considerations and possible updates.

In the previous section, we devoted our attention to some details concerning some individual teams in different leagues. It became clear throughout those analysis that a continuous time scoring system would increase the pressure on the teams in order that each time of the match may be important to the final classification, and transfer the importance of the final result to the all part of the game.

Concerning the competitiveness of the overall league, some indicators are presented on Table 5. There we may see that, generally, the gaps between different positions decrease in very significant way. In fact, the difference between the points held by the teams classified in different positions w.r.t. the champion decreases in almost all the scenarios when compared with the traditional 3-for a win system. The indicators $1^{st}$-$3^{rd}$, $1^{st}$-$9^{th}$ and $1^{st}$- last classified, always reduce the classification gap between those competitors, except for the top 3 on Spanish league in 2015-2016.

The number of teams that reach the leadership somewhere over the season is increased in almost all the simulated seasons. Another way to measure the competitiveness is by the number of swaps in leadership, and those are significantly increased with new scoring system. Finally, a small side-note to remark that this system turns highly improbable to have draws in the final classification of a league, as the final points are result of a continuous scale, and not a sum of integer values.

| Indicators | Premier 15-16 | | La Liga 15-16 | | NOS 15-16 | | NOS 16-17 | |
|---|---|---|---|---|---|---|---|---|
| | New | Actual | New | Actual | New | Actual | New | Actual |
| Gap 1st-3rd place (%) | 4,8 | 13,6 | 7,0 | 3,3 | 14,2 | 17,0 | 11,1 | 14,6 |
| Gap 1st-9th place (%) | 15,9 | 37,0 | 38,2 | 47,3 | 37,1 | 53,4 | 33,1 | 47,6 |
| Gap 1st-last (%) | 56,0 | 79,0 | 54,7 | 64,8 | 57,9 | 71,6 | 55,3 | 74,4 |
| # Overall Changes | 424 | 400 | 407 | 422 | 341 | 307 | 315 | 277 |
| # Changes on Leadership | 15 | 5 | 8 | 6 | 11 | 8 | 9 | 2 |
| # Different Leaders | 6 | 4 | 5 | 4 | 4 | 4 | 4 | 3 |
| Aver. points per game | 1,25 | 1,36 | 1,27 | 1,38 | 1,26 | 1,38 | 1,26 | 1,37 |

Table 5: Overall indicators for the new proposal and present scoring system.

In a strategy point of view, this continuous scoring system presents, in our opinion, several advantages and may, if adopted, introduce major changes in this sport: it will make the teams to go forward for attack since the beginning of the game, as the main objective is not to finish the match scoring more goals than the opponent (the trinary point of view) but to be winning as much part of the match as it may be possible. Also, the interest in the game may be better, at a time when statistics and data regarding the game are becoming more widespread and interesting by fans, especially the younger ones who use the same data to make entertainment decisions in skill sports which are in great expansion. The social impact of the scoring mix would also have an even more dynamic social effect than actual trinary scheme, since the discussion between supporters about the referee's decision, the failed penalty or the ball in the post at a given minute would affect the classification of a given team or even the decision of the winner of the championship of a certain season. In another point of view, even the most defending teams who, so many times, still defend when losing by one goal to try the recover at the end of the game must to rethink its own strategy, as a goal in the end of the match that gives a draw or even a victory in the final seconds of the match won't be so important as the final points will be acquainted in terms of the proportion of time where the match was in their favour (winning or at draw). Finally, from the technical and tactical point of view, all of this would oblige the coaches to change their models of approach to the game.

On the other hand, a game where a team is winning since the early minutes of the match, may lose some competitiveness in the late part of the game as the points in dispute are fewer as the match is near its end. Also, the "early-in-advantage" team may be tempted to adopt a defending strategy since that early time, resulting on a lack of attractiveness for the remaining of the game. According to this point of view, there are several points of view that may be exploited. For example, a mixed strategy between the 3-for-a-win and the new proposal may be built, as the following formula w.r.t. the team A match final points represents,

$$P_A = \frac{1}{2} * \frac{3T_{WA} + T_{dA}}{T_m} + \frac{1}{2} * FR_A$$

where $FR_A$ takes the value 3,1 or 0 if the team (A) wins, draw or loose the match, respectively. As half of the match points are attributed by the new system and the other half by the final result as nowadays, this balances the interest of the game through all the match length as in the final minutes of the game there are still a significant number of points to distribute. This scenario is easy to simulate based on the tables presented on previous section, as the final points will be just numerically equal to the averages between the two systems final scorings, but in order to keep this paper's length acceptable, we postpone the description of these results to future work.

Another option is to take into account the victory gap. In fact, with any of the systems here presented, a team which is winning by 1-0, doesn't have any immediate incentive to look for a greater advantage, excepting increasing the probability of final victory.

However, this fact is somehow faded by the new system because as long as the time passes less match points are in dispute. As so, we updated this system taking also into account, in the final result, the goals difference between the two teams. This idea can be translated into the following formula:

$$P_A = \frac{1}{3} * \frac{3T_{WA} + T_{dA}}{T_m} + \frac{1}{3} * FR_A + \frac{1}{3} * GD_A$$

where $GD_A$ remains for some function that translates the goals difference between team A and its opponent into the final result. For example, $GD_A$ may be set to 0,1,2 or 3 if the goals difference between team A scored and against is ≤0,1,2 or ≥3 goals, respectively. To the best of our knowledge, the scoring ideas presented on this paper had never been applied, tested or even reported. As so, the main goal for this article, which was to bring new ideas to the community, is accomplished allowing that new work may be done by the scientific community, which may use several branches of science to discuss the validity of these ideas.

## 5. Conclusions

In this paper, we present a new time-dependent scoring system. Several tests were made using real data from Premier League, La Liga and Liga NOS. Based on those datasets and in some fundamental analysis we conclude that the proposed system reduces the relative gaps between the teams, when compared with the traditional 3-for-a-win scoring methodology. If applied in real competitions, this could lead to more competitiveness during the leagues with significant more position swaps among the different competitors. Besides that, a discussion about some other consequences of this methodology is made, to stress out benefits and disadvantages, proposing some ideas to tackle the latter ones.

**Funding:** This research did not receive any specific grant from funding agencies in the public, commercial, or not-for-profit sectors.

# Bibliography

Boor, S., Hanson, C., & Ross, C. (2018). *Deloitte Football Money League 2018.* UK: Deloitte.

Daily Mail. (2017, Jan 12). *Manchester United have most visited football website in the world with Arsenal and Liverpool close behind as Premier League leads the way*. Retrieved from Mail online: http://www.dailymail.co.uk/sport/football/article-4110822/Manchester-United-visited-football-website-world-Arsenal-Liverpool-close-behind.html

Fernandez-Cantelli, E., & Meeden, G. (2003). An Improved Award System for Soccer. *CHANCE*, 23-29.

FIFA. (2007, May 31). *FIFA Big Count 2006: 270 million people active in football*. Retrieved from FIFA.com: http://www.fifa.com/media/news/y=2007/m=5/news=fifa-big-count-2006-270-million-people-active-football-529882.html

Haugen, K. K. (2007). Point Score Systems and Competitive Imbalance in Professional Soccer. *Journal of Sports Economics*.

Hellier, D., Penty, R., & Mayes, J. (2017, Oct 12). *Premier League Viewing Recovers While Facebook and Amazon Loom.* Retrieved from Bloomberg.com: https://www.bloomberg.com/news/articles/2017-10-12/premier-league-soccer-gains-viewers-in-season-start-on-sky-bt

La Liga. (2016, Oct 01). *Historical Stats*. Retrieved from La Liga: http://www.laliga.es/en/statistics-historical/calendar/primera/2015-16/jornada-01/

Liga Portugal. (2017, Jul 01). *Liga NOS*. Retrieved from Liga NOS: http://liganos.pt/

Premier League. (2016, Oct 01). *Tables*. Retrieved from PremierLeague.com: https://www.premierleague.com/tables?co=1&se=42&ha=-1

ZOS, Lda. (2016, Jul 01). *Campeonato Espanhol 2015/16*. Retrieved from ZeroZero.pt: http://www.zerozero.pt/edition.php?id=87618